\documentclass[12pt,preprint]{aastex}

\slugcomment{Last Modified \today}

\shorttitle{HI Gas in Nearby AGN-Hosting Galaxies}
\shortauthors{Y.-N. Zhu & H. Wu}

\begin{document}

\title{A Statistic Study of HI Gas in Nearby Narrow Line AGN-Hosting Galaxies}
\author{Yi-Nan Zhu\altaffilmark{1}, Hong Wu\altaffilmark{1}}
\altaffiltext{1}{Key Laboratory of Optical Astronomy, National Astronomical Observatories, Chinese Academy of Sciences, Beijing 100012, China; zyn@bao.ac.cn; hwu@bao.ac.cn}

\begin{abstract}

As a quenching mechanism, AGN feedback could suppress on-going star formation in their host galaxies. On the basis of a sample 
of galaxies selected from ALFALFA HI survey, the dependence of their HI mass (M$_{HI}$), stellar mass (M$_{*}$) \& HI-to-stellar 
mass ratio (M$_{HI}$/M$_{*}$) on various tracers of AGN activity are presented and analyzed in this paper. Almost all the 
AGN-hostings in this sample are gas-rich galaxies, and there is no any evidence to be shown to indicate that the AGN activity 
could increase/decrease either M$_{HI}$ or M$_{HI}$/M$_{*}$. The cold neutral gas can not be fixed positions accurately just 
based on available HI data due to the large beam size of ALFALFA survey. In addition, even though AGN-hostings are more easily 
detected by HI survey compared with absorption line galaxies, these two types of galaxies show similar star formation history. 
If an AGN-hosting would ultimately evolve into an old red galaxy with few cold gas, then when and how the gas has been exhausted 
have to be solved by future hypotheses and observations.

\end{abstract}

\keywords {galaxies: evolution --- galaxies: active --- radio lines: galaxies --- galaxies: nuclei}

\section{Introduction}

Since the discovery of active galactic nuclei (AGNs) seventy years ago, there has been a rapid advance in our understanding 
of them. Recently, one important progress in the study of AGNs is that there are accumulating evidences supporting the co-evolution 
scenario of AGNs and their host galaxies (see the review by \citet{kormendy13}), and this co-evolution has been supported 
by several observational facts \citep[e.g.,][]{magorrian98,tremaine02,cowie03,greene06,woo10}. For nearby galaxies, a number 
of spectroscopic surveys of nuclei have been conducted, and these surveys show a fact that the galaxies hosting weak AGNs 
are very popular in the local universe \citep{ho97}. Therefore, the investigation of the interaction between AGNs and their 
host galaxies is essential for understanding the evolution of galaxies.

In order to describe the co-evolution of AGNs and galaxies, many models have involved thea AGN feedback \citep[e.g.,][]{fabian99, granato04, hopkins06}: 
the energetic output from AGN heat surrounding medium (both gas and dust) and expel them away. The dense gas is not only vital 
to the growth of central black hole, but also indispensable as the raw material to form stars \citep{kennicutt98a, kennicutt12}, 
hence the feedback from AGNs could suppress surrouding star formation.

Hydrogen is the most abundant element in the universe, so neutral atomic hydrogen (HI) is a frequently used tracer of cold gas. 
As the raw material to feed the central nuclei and to form new stars, HI plays a fundamental role in the formation and evolution 
of galaxies. But HI has not direct connection with current star formation. HI need to transform to molecular hydrogen, then 
the dense core embeded in molecular cloud condense enough to evolve into protostars \citep{bergin07}. However, observational 
information of the HI could still provide crucial constraints on the star formation theory: the surface density of HI has 
been firstly scaled to the surface density of star formation rate by \citet{schmidt59}, then has been improved by \citet{kennicutt89, kennicutt94, kennicutt98b}; 
in the last decade, many studies, such as \citet{evoli11,toribio11a,toribio11b,huang12}, have investigated how the HI content 
varies with stellar properties in normal galaxies.

AGN feedback could heat the gas reservoir that have been built up before the AGN activity began, or even drive part of the gas out of its host \citep[e.g.,][]{dimatteo05}.
In order to investigate the interaction between the active black hole and surrounding gas, the deep and large area blind HI 
survey is necessary \citep{heckman78, mirabel84}. \citet{konig09} showed a study of HI in 27 nearby low-luminosity quasi-stellar 
objects (QSOs). The median HI mass in their whole sample is of the order of 11.4$\times$10$^{9}$ $M_{\rm\odot}$, which is 
a factor of two higher than the HI content of milky way. This means that HI in these low-luminosity QSOs is abundant. However, 
they did not find strong correlation between HI mass and IR luminosity, though the IR luminosity is usually considered to be 
connecting with the intensity of star formation and AGN \citep[e.g.,][]{kennicutt98a, sanders96}. \citet{ho08a} performed 
a HI emission survey for more than one hundreds local galaxies, and all of these galaxies host broad-line (type-I) AGN in 
their centers. They found that the host galaxies of type-I AGNs, including those ones which are luminous enough to be qualified 
as quasars, were generally gas-rich systems such as the quiescent objects of the same morphological type \citep{ho08b}. 
On the basis of the data of Herschel Reference Survey \citep[HRS;][]{boselli10}, which is a combination of multi-wavelength 
observations of more than 300 local galaxies, \citet{hughes09} did not find any connection between the AGN feedback and the 
decrease of the amount of neutral gas; on the contrary, they proposed that the environmental effects might play a significant 
role in the suppressing of star formation.

According to the AGN unified scheme \citep{antonucci93, urry95}, the narrow-line (type-II) AGN is those to be observed through 
an obscuring structure, such as torus \citep{hatziminaoglou09}, which prevents a direct view of the emission lines from broad-line 
regions. Therefore, for the narrow-line AGNs, the effect of AGN feedback acting on gas content in host galaxies should also 
be investigated. Ordinarily, narrow-line AGNs are selected on the basis of the widths and flux ratios of some optical emission 
lines. Based on this selection methodology and stacking technique for measuring the average HI content, \citet{fabello11} 
conducted an investigation of a sample of narrow-line AGNs from Arecibo Legacy Fast ALFA survey \citep[ALFALFA;][]{giovanelli05a,giovanelli05b,giovanelli07,haynes07}.
\citet{fabello11} did not find significant difference in HI content between galaxies with and without AGNs. However, they did 
not discriminate the Seyfert galaxies \citep{seyfert43} from LINERs \citep[stand for Low-ionization nuclear emission-line regions;][]{heckman80}, 
and still contained composite galaxies defined by \citet{kewley06} as well.

In this work, an observational study of effect of strength of narrow-line AGNs on neutral gas located in their host galaxies 
will be performed, by utilizing the observations of some large area surveys released recently. The structure of the paper is 
as follows. The construction of the sample and the estimation of AGN activity are described in $\S$2. 
The major results are presented in $\S$3. Some discussions and a summary of this work are given in $\S$4 and $\S$5. 
Throughout this paper, a $\Lambda$CDM cosmology with $\Omega_{\rm m}=0.3$, $\Omega_{\rm \Lambda}=0.7$ and $H_{\rm 0}=70\,{\rm km \, s^{-1} Mpc^{-1}}$ was adopted.

\section{Sample and Data Reduction}

\subsection{ALFALFA Survey}
As the most popular second generation blind HI survey, the ALFALFA survey initiated in 2005 Feb, and it would map about 7000
deg$^{2}$ of high galactic latitude northern sky with declination lower than +36 $\deg$. As the largest blind HI survey 
so far, when complete, ALFALFA would detect more than 30,000 galaxies. This survey used a seven-beam array named Arecibo L-band 
feed array (ALFA) mounted on the Arecibo 305m antenna with the spectral resolution of ~5.5km/s and beam size of ~3.5$\arcmin$, 
that is about 4 times better than that of previous HI surveys, such as HIPASS \citep[stand for HI Parkes All Sky Survey;][]{barnes01, meyer04}. 
Considering the excellent sensitivity of Arecibo antenna, ALFALFA could detected more deep universe with boundary of about 250Mpc.

Recently, A pilot HI catalog has been released to public by ALFALFA team \citep{haynes11,martin10}. The coverage of the pilot 
HI survey is about 2800 deg$^{2}$ of sky, which is about 40\% compared to final survey area. Hence, the released pilot HI 
catalog, containing the flux, mass and line width of HI, was called as $\alpha$.40 catalog. This catalog contains 15855 sources 
(15041 of them are extra-galactic HI sources) in four parted regions, two in the spring northern sky and two in the fall northern 
sky. In order to match with optical spectral data, only the spring sources were selected in this paper 
(130.0$\arcdeg$$<$RA$<$240.0$\arcdeg$ \& 4.0$\arcdeg$$<$DEC$<$16.0$\arcdeg$, 120.0$\arcdeg$$<$RA$<$247.5$\arcdeg$ \& 24.0$\arcdeg$$<$DEC$<$28.0$\arcdeg$). 
The ALFALFA survey detects HI sources out to distance of about 250Mpc (corresponding to redshift z$\sim$0.06), but a gap 
around 225Mpc is exist due to the disturbance from the radar located at a nearby airport. In order to avoid this contamination, 
a redshift range of 0.01$<$z$<$0.05 was used in this work. Here, the setting of lower redshift limit is for the sake of adequate 
cover area to perform classification based on the optical spectra. After that, there are totally 7019 HI sources left.

\subsection{The Sloan Digital Sky Survey}
The optical spectral sample galaxies are selected from the Sloan Digital Sky Survey \citep[SDSS;][]{york00}. A series of processed 
catalogs containing data of SDSS release seven \citep[DR7;][]{abazajian09} supplied in MPA website 
\footnote{http://www.mpa-garching.mpg.de/SDSS/DR7/} were downloaded. A total of $927,552$ sources 
are listed in these catalogs. Except raw data, some derived physical quantities, such as spectral redshift, the fluxes of 
optical emission lines, stellar masses \& metallicities, are supplied as well. The detailed descriptions about the data reductions 
could been found in a series of papers \citep{kauffmann03a, tremonti04, salim07}. The spectral observation of SDSS is deeper 
than ALFALFA survey. In the two spring regions listed above with the redshift range of 0.01$<$z$<$0.05, there are 23009 SDSS 
optical spectral sources. 

The SDSS spectral sample were matched with the 7019 HI sources with the radius of 3$\arcsec$, and 6762 of them have been included 
in $\alpha$.40 catalog. There is not any galaxies with multiple matches with the radius of 3$\arcsec$. If we enlarged the 
matching radius to 5, 7, 10, 15$\arcsec$, there would be 6887, 6926, 6953, 6972 HI sources left after cross-matching. Hence, 
for the HI sources, the completeness of our sample is about 97\%. This cross-matched sample (with the radius of 3$\arcsec$) 
containing optical and HI observations will be called as $\alpha$.40-SDSS sample throughout this paper.  

\subsection{Spectral Classification}

In order to select type-II AGNs, first of all, the emission lines galaxies should be separated from the absorption line ones. 
Traditionally, the H$\alpha$ equivalent width EW(H$\alpha$)$\sim$0.0\AA{}, which means only continuum could be detected around 
rest-frame 6563\AA{}, was considered as the border of emission and absorption line galaxies \citep{zhu10}. But in this work, 
for avoiding cross-contamination between emission and absorption line galaxies, EW(H$\alpha$)$\sim$-1.0\AA{} is used as the 
upper limit to choose emission line galaxies. For the derived spectral lines in MPA-JHU catalogs, the positive and negative 
signs means absorption and emission, respectively. These definition would also be employed in this study: the galaxies with 
EW(H$\alpha$)$>$1.0\AA{} are classified as absorption line galaxies; those with the absolute value, $|$EW(H$\alpha$)$|$$<$1.0\AA, 
are treated as transition galaxies. 

\begin{figure}
\figurenum{1}
\epsscale{1}
\plotone{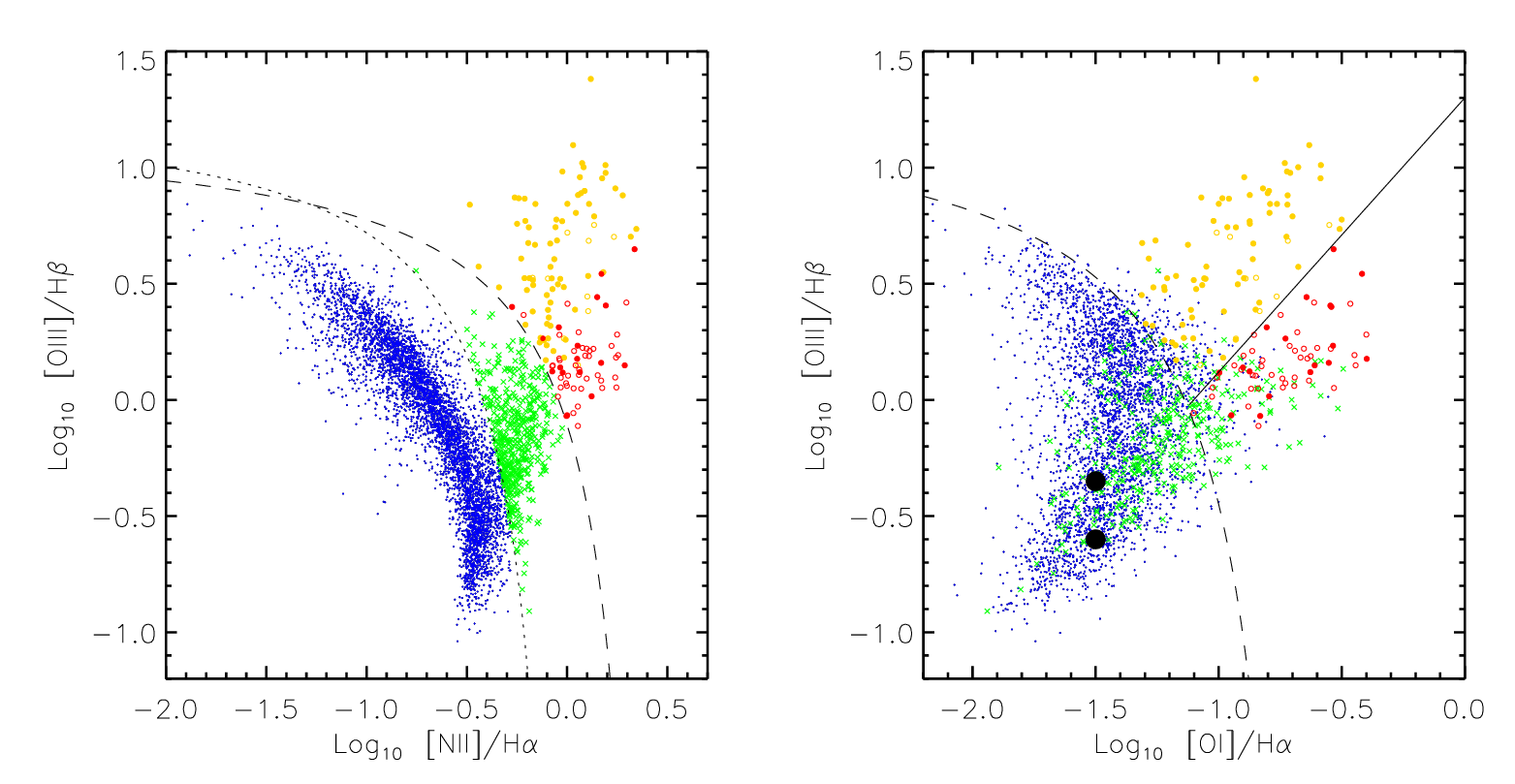}
\caption{The BPT diagnostic diagram: [NII]/H$\alpha$ vs [OIII]/H$\beta$ \& [OI]/H$\alpha$ vs [OIII]/H$\beta$. In the left 
panel, the criteria from \citet{kauffmann03a} and \citet{kewley01} are illustrated as dotted and dashed curves, respectively. 
The objects below the dotted curve, represented by blue dots, are defined as star-forming galaxies. 
The green crosses between the above two curves are those classified as composite galaxies.
The open and solid circles above the dashed curve denote AGNs, the circles with red and orange colors represent LINERs \& Seyferts, respectively. 
In the right panel, the dashed curve is the criteria from \citet{kewley01}. Blue dots and green crosses denote star-forming 
and composite galaxies classified in the left panel. A solid line raising from the dashed curve, defined by \citet{kewley06}, 
is a boundary to distinguish Seyferts (orange) from LINERs (red). The solid circles above the dashed curve represent the Seyferts or 
LINERs with EW(H$\alpha$)$<$-3.0\AA{}; the open circles represent the ones with -3.0\AA$<$EW(H$\alpha$)$<$-1.0\AA. The two big black solid circles in the right panel denote the base points for Seyfert and LINER branches to estimate 
the distance from the star-forming sequence d$_{SF}$ defined by \citet{kewley06}.
}
\label{fig1}
\end{figure}

For the emission line galaxies, the optical spectral classifications of galaxies were made by adopting the traditional 
BPT diagnostic diagram: [NII]/H$\alpha$ versus [OIII]/H$\beta$ \citep{baldwin81, veilleux87}, as shown in Figure~\ref{fig1}. 
Only the sources with the signal-to-noise ratio (S/N) of H$\alpha$, H$\beta$, [OIII], [NII] emission lines greater than 3$\sigma$ were selected.
The dashed curve in the left panel of Figure~\ref{fig1} is from \citet{kauffmann03a} (hereafter Ka03 line) and the dotted curve is 
from \citet{kewley01} (hereafter Ke01 line). The objects located below the Ka03 line were classified as star-forming galaxies; 
those above the Ke01 line were classified as narrow-line AGNs; those between the above two lines were classified as composite 
(starburst $+$ AGN) galaxies \citep{kewley06,wu98}. For comparison, the total 23009 SDSS sources were also classified in the 
same way. Table 1 lists the numbers of galaxies after spectral classifying. 

The sources that were classified as narrow-line AGNs include type-II Seyferts and LINERs. However, it is difficult to distinguish 
type-II Seyferts from LINERs simply based on the [NII]-to-H$\alpha$ flux ratio, because this flux ratio is not sensitive to 
the hardness of the ionizing radiation field. \citet{kewley06} have taken advantage of the other two BPT diagnostic diagrams 
([SII]/H$\alpha$ vs. [OIII]/H$\beta$ and [OI]/H$\alpha$ vs. [OIII]/H$\beta$) to seperate Seyferts and LINERs. In this study, 
only the [OI]/H$\alpha$ vs. [OIII]/H$\beta$ diagram is used. d$_{SF}$ is a parameter to represent the distance from the 
star-forming sequence (next subsection) based on the flux ratio of [OI]-to-H$\alpha$ and [OIII]-to-H$\beta$. Only the galaxies 
have the measured S/N of [OI] emission line greater than 3$\sigma$ is used to distinguish type-II Seyferts from LINERs. For 
the 153 AGNs in $\alpha$.40-SDSS sample described above, 141 of them satisfy this criterion.

The emission-line fluxes of [OI] \& H$\alpha$ have been corrected for foreground and intrinsic extinction because of the relatively 
large wavelength gap of $\sim$250\AA{} between [OI] \& H$\alpha$. The foreground extinction was corrected by assuming the \citet{cardelli89}'s 
extinction curve and R$_{\rm V}$=3.1, and the intrinsic extinction correction was corrected based on the color excess E(B-V) 
which was acquired from the Balmer decrement $F_{\rm H\alpha}/F_{\rm H\beta}$ \citep{calzetti01}. \citet{kewley01} presented 
maximum starburst line on every BPT diagnostic diagrams, which could be seen as the upper limit of emission line ratios ionized 
by photons from young hot stars. After that, 3 sources were deleted from AGNs: these 3 sources were the ones to be classified 
as AGNs in the [NII]/H$\alpha$ vs. [OIII]/H$\beta$ diagram but below the upper boundary (dotted black curve in the right panel 
of Figure~\ref{fig1}) of the star-forming galaxies in the [OI]/H$\alpha$ vs. [OIII]/H$\beta$ diagram. In the right panel of 
Figure~\ref{fig1}, the solid line raised from the dotted curve are the boundary to separate Seyferts from LINERs defined by 
\citet{kewley06}: the upper branch are classified as Seyferts (orange circles), and the lower branch are classified as LINERs (
red circles). There are totally 80 Seyferts and 58 LINERs in $\alpha$.40-SDSS sample. In the right panel of Figure~\ref{fig1}, 
the star-forming galaxies and composite galaxies classified in [NII]/H$\alpha$ vs. [OIII]/H$\beta$ diagram are also demonstrated 
with blue dots and green crosses respectively.

Some studies, such as \citet{capetti11a}, proposed that traditional BPT diagnostic diagrams were not a useful tool to reveal 
the distinction among various emission line galaxies, becasue the equivalent widths of these optical lines were not involved 
in traditional BPT diagnostic. \citet{cid11} introduced the criterion of EW(H$\alpha$)$<$-3.0\AA{} as a necessary condition 
for the existence of active nucleus. They proposed that the AGNs selected based on traditional BPT diagrams with EW(H$\alpha$)$>$-3.0\AA{} 
are retired galaxies \citep{cid10} who might have a fake AGN located in the central region of the hosts. The output of Balmer 
emission lines of retired galaxies could be from the photoionization by stellar populations older than 100 million years. 
In this work, the AGNs (both Seyferts and LINERs) with EW(H$\alpha$)$<$-3.0\AA{} are distinguished with the others, and numbers 
of them are 70 \& 18 for Seyferts \& LINERs, respectively (see Table 2.). In both panels of Figure~\ref{fig1}, the solid orange 
\& red circles  represent the Seyferts and LINERs with EW(H$\alpha$)$<$-3.0\AA{}, and the open orange \& red circles represent the ones with -3.0\AA$<$EW(H$\alpha$)$<$-1.0\AA. 

\subsection{AGN activity}
For a galaxy whose output is dominated by its central AGN, its emission of [OIII] is considered to be originated from narrow 
line region and excited by central AGN. The luminosity of the [OIII]$\lambda$5007 emission line (hereafter as L$_{[OIII]}$) 
are merely suspected to scale with the AGN bolometric luminosity \citep{heckman04}, so L$_{[OIII]}$ is statistically used 
as a tracer of AGN activity for narrow-line AGNs by many authors \citep{risaliti99,kauffmann03a,wang08,wang13}. The optical 
[OIII] emissions suffer from the dust extinctions from both Milky Way and the host galaxy. In this study, the foreground Galactic 
extinction was first corrected by assuming the \citet{cardelli89}'s extinction curve and R$_{\rm V}$=3.1. Then the intrinsic 
extinction correction was performed, based on the color excess E(B-V) which was acquired from the Balmer decrement $F_{\rm H\alpha}/F_{\rm H\beta}$ \citep{calzetti01}.

The ionizing radiation fields related with AGNs are often harder than the ones excited by young stars. 
For Seyfert and LINER branches respectively, \citet{kewley06} presumed two base points on the [OI]/H$\alpha$ vs. [OIII]/H$\beta$ 
diagram (the two big black solid circles in the right panel of Figure~\ref{fig1}).
For a Seyfert or LINER, the d$_{SF}$ is the normalized distance to the corresponding base point. The normalization, not a 
special physical quantity but just for convenience to compare between this two types of galaxies, was performed by dividing 
1.55 dex or 1.65 dex for Seyferts or LINERs. Compared with those with smaller d$_{SF}$, the AGNs with larger d$_{SF}$ could 
be considered that the percentage of the contributions from star formations are much smaller in the total output of the whole 
galaxies \citep{wu07}. Moreover, the ratios of [OIII]-to-H$\beta$ and [OI]-to-H$\alpha$ are the projections of d$_{SF}$: the 
ratio of [OIII]-to-H$\beta$ are sensitive to the hardness of radiation field; while the ratio of [OI]-to-H$\alpha$ has been 
found roughly correlated with the X-ray spectral slopes, which should be effected by the power of AGN \citep{wang10}. For 
Seyferts, \citet{kewley06} have illustrated the d$_{SF}$ increased as a function of L$_{[OIII]}$, the traditional tracer of 
AGN activity described above. But for LINERs, the relation between d$_{SF}$ and L$_{[OIII]}$ is unclear. 

\section{Results}

\begin{figure}
\figurenum{2}
\epsscale{1}
\plotone{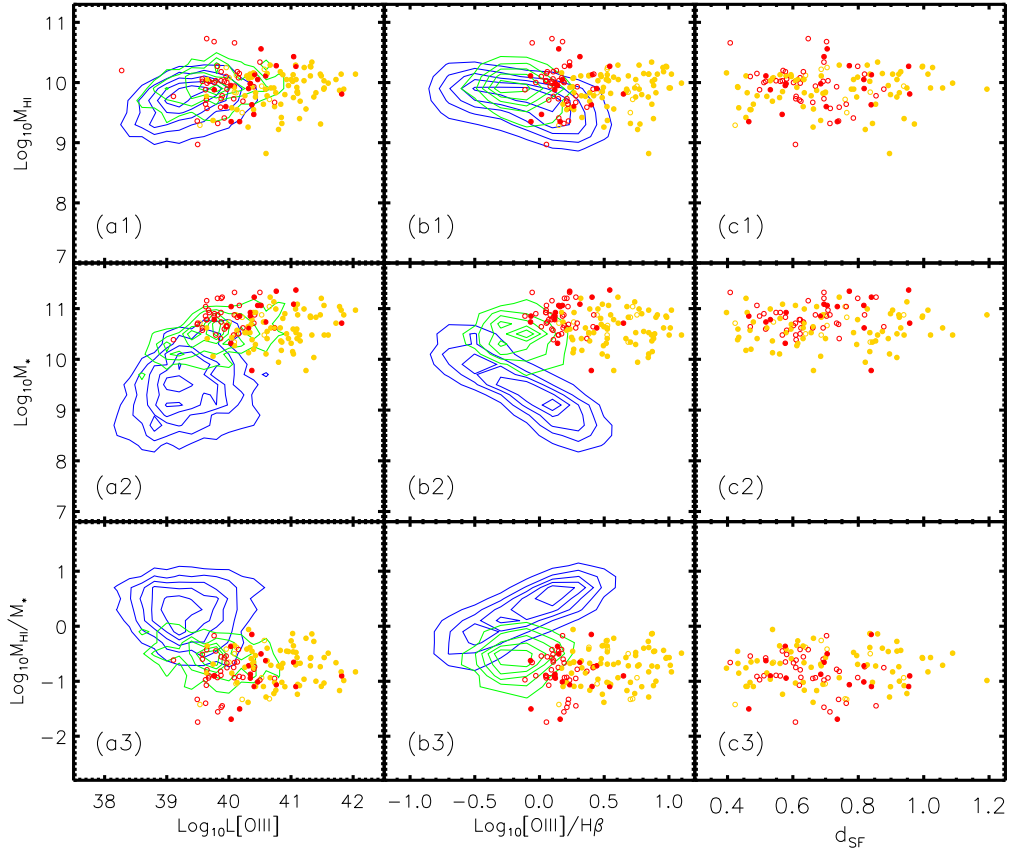}
\caption{Comparing the M$_{HI}$, M$_{*}$ \& M$_{HI}$/M$_{*}$ with various factors, including L$_{[OIII]}$ (a1, a2, a3), 
[OIII]/H$\beta$ (b1, b2, b3), d$_{SF}$ (c1, c2, c3). The solid orange \& red circles in these panels represent the Seyferts 
and LINERs with EW(H$\alpha$)$<$-3.0\AA{}; the open orange \& red circles represent the Seyferts and LINERs with 
-3.0\AA$<$EW(H$\alpha$)$<$-1.0\AA. Blue and green contours denote the locations of star-forming galaxies and composites, respectively.
}
\label{fig2}
\end{figure}

Figure~\ref{fig2} shows the variations of M$_{HI}$ (HI mass), M$_{*}$ (stellar mass) \& M$_{HI}$/M$_{*}$ (HI-to-stellar mass ratio) 
with some factors tracing AGN activity: L$_{[OIII]}$ (a1, a2, a3), the ratio of [OIII]-to-H$\beta$ (b1, b2, b3) \& d$_{SF}$ (c1, c2, c3). 
The M$_{*}$ was derived by fitting optical photometries of SDSS with population synthesis models \citep{bruzual03}, and was 
provided by MPA-JHU in their catalogs. The solid orange \& red circles in Figure~\ref{fig2} are the Seyferts and LINERs with 
EW(H$\alpha$)$<$-3.0\AA; the open orange \& red circles are the Seyferts and LINERs with -3.0\AA$<$EW(H$\alpha$)$<$-1.0\AA. 
For comparison, the locations of star-forming and composite galaxies are shown with blue and Green contours respectively. 
Since no d$_{SF}$ could be estimated to quantify the AGN activity for composite galaxies, the ratio of [OIII]-to-H$\beta$ 
is employed as the representative of d$_{SF}$ in the middle column panels of Figure~\ref{fig2}. The ratio of [OIII]-to-H$\beta$ 
could be an indicator of radiation field hardness\citep{hogg05} and be simply seen as the projection of d$_{SF}$. In addition, 
it must be noted that for most of the low redshift star-forming and composite galaxies, their L$_{[OIII]}$ might be underestimated, 
because the most outputs of [OIII] in these galaxies are from star formation regions, which can spread all over the galaxy, 
instead of the central regions covered by the 3$\arcsec$ diameter fibers of SDSS.

In Figure~\ref{fig2}, we can see that most of the galaxies, including star-formings, composites \& AGNs, show the similar M$_{HI}$ 
range from 10$^9$M$_{\odot}$ to 3$\times$10$^{10}$M$_{\odot}$. Star-forming galaxies have wide ranges of both M$_{*}$ \& M$_{HI}$/M$_{*}$. 
Their M$_{*}$ ranges from 10$^8$M$_{\odot}$ to 10$^{11}$M$_{\odot}$. But the galaxy with M$_{*}$ lower than 5$\times$10$^9$M$_{\odot}$ 
is definitely absent in other types of galaxies. For the star-forming galaxies with M$_{*}$ lower than 5$\times$10$^{10}$M$_{\odot}$, 
their [OIII]-to-H$\beta$ are obviously higher than the other star-formings. They are supposed to be in the upper-left star-forming 
branch in [NII]/H$\alpha$ versus [OIII]/H$\beta$ diagnostic diagram with lower metallicity \citep{wu07}. Additionally, a large 
amount of the composite galaxies, represented by green contours in Figure~\ref{fig2}, show signs of sufficiency in both the 
M$_{HI}$ and M$_{*}$, just like the AGN-hostings rather than most of star-formings.

For the AGN-hosting galaxies (both Seyferts and LINERs) presented in Figure~\ref{fig2}, some characters could be seen: first 
of all, approximately all the AGN-hosting galaxies (both Seyferts and LINERs) in $\alpha$.40-SDSS sample are gas-rich galaxies, 
with the M$_{HI}$ greater than 2$\times$10$^9$M$_{\odot}$, which indicates the neutral gas content in AGN-hostings is as abundant 
as in star-forming and composite galaxies; Secondly, besides neutral gas, the AGN-hostings also possess a large number of 
stars; Thirdly, LINERs in $\alpha$.40-SDSS sample prefer to have larger M$_{*}$ than Seyferts: aside from few ones, the M$_{*}$ 
in almost all the LINERs are greater than 3$\times$10$^{10}$M$_{\odot}$, but this value can not be simply considered as the 
lower M$_{*}$ limit for Seyferts in this sample; Fourthly, AGN-hosting have plenty of neutral gas and stars, which leads to 
the lower M$_{HI}$/M$_{*}$ compared with a lot of star-formings; Finally and most important of all, for the AGN-hosting galaxies 
(both Seyferts and LINERs) in $\alpha$.40-SDSS sample, there is no evidence to indicate the dependence of M$_{HI}$, M$_{*}$ 
\& M$_{HI}$/M$_{*}$ on the AGN activity: both M$_{HI}$ and M$_{*}$ do not increase or decrease with the raising of either 
L$_{[OIII]}$ or d$_{SF}$.

\section{Discussion}

\subsection{Aperture Effect}

The aperture of the SDSS fiber is 3$\arcsec$. For the galaxy hosting active nucleus in its center, its emission lines are 
usually from a confined small region. As an example, \citet{ho97} detected ionized gas in the central few hundred parsecs 
of a large number of local galaxies to search active nuclei. But if the emitters of nebular lines spread all over the galaxies 
rather than the central part, such as most of late-type spiral galaxies do, the aperture effect could not be ignored \citep{moran02, maragkoudakis14}, 
and an aperture correction has to be done to derive the total luminosities of these emission lines \citep[e.g.,][]{hopkins03}. 

The galaxies in this work cover a wide redshift range from $0.01$ to $0.05$, and the 3$\arcsec$ diameter of SDSS fibers corresponds
to a large range of physical size ($0.6$, $1.2$, $1.8$, $2.4$, $2.9$kpc at redshift z$\sim$$0.01$, $0.02$, $0.03$, $0.04$, $0.05$).
Hence, for the area covered by the fiber's aperture, there is a remarkable variation of about $23$ when the redshift increase 
from $0.01$ to $0.05$. As described above, two parameters, L$_{[OIII]}$ \& d$_{SF}$ are used to trace AGN activity. For d$_{SF}$, 
which is derived from ratios of some emission lines, the aperture effect should not be taken into account by assuming the 
spatial coincidence of the emitters of these lines. Nevertheless, for L$_{[OIII]}$, if the [OIII] emitters do not concentrate 
in the central regions of the galaxies, the underestimation of L$_{[OIII]}$ would be inevitable.

\begin{figure}
\figurenum{3}
\epsscale{1}
\plotone{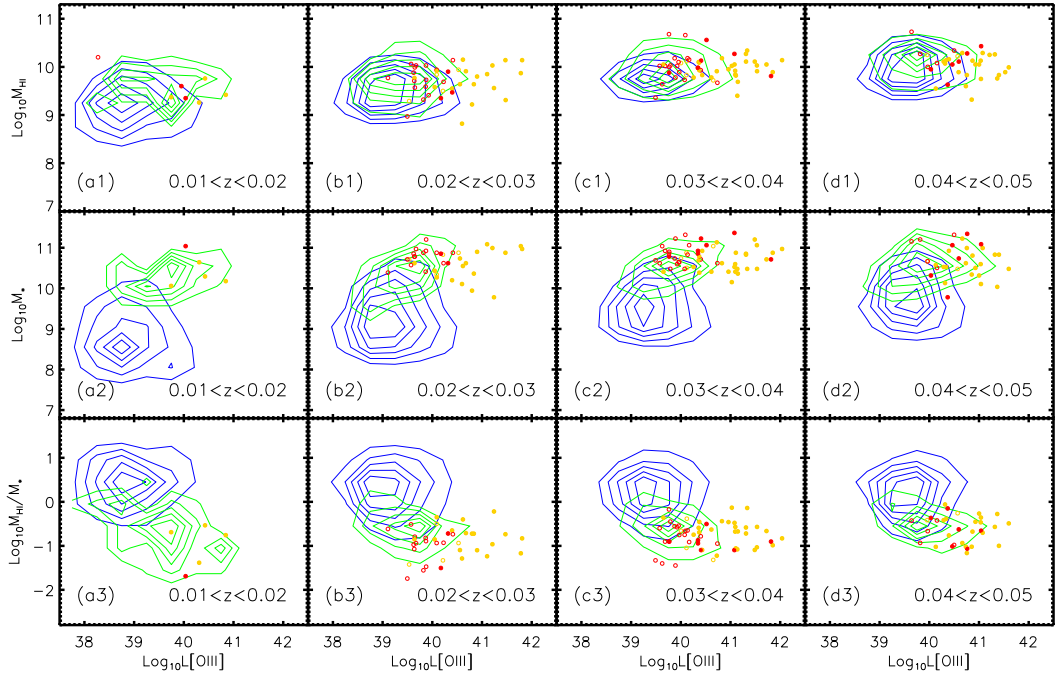}
\caption{Comparing the M$_{HI}$, M$_{*}$ \& M$_{HI}$/M$_{*}$ with L$_{[OIII]}$ in four different redshift bins. 
From left to right, the redshift coverage is: 0.01$\sim$0.02, 0.02$\sim$0.03, 0.03$\sim$0.04, 0.04$\sim$0.05. 
The denotations of the symbols and contours are the same as in Figure~\ref{fig2}.
}
\label{fig3}
\end{figure}

Will the aperture affect the results obtained above of correlations between HI abundance and AGN activity quantified by L$_{[OIII]}$? 
In order to examine it, the galaxies in $\alpha$.40-SDSS sample are grouped into several redshift bins (0.01$\sim$0.02, 0.02$\sim$0.03, 
0.03$\sim$0.04, 0.04$\sim$0.05), as shown in Figure~\ref{fig3}. We could find that, the panel columns from left to right in 
Figure~\ref{fig3}, both the M$_{HI}$ \& M$_{*}$ increase slowly with the raising of redshift for all types of galaxies. For 
star forming galaxies, which constitute the largest sub-sample in $\alpha$.40 sample, their median values of $log_{10}$M$_{HI}$ 
and $log_{10}$M$_{*}$ in the unit of solar mass are 9.18, 9.49, 9.75, 9.96 and 8.66, 9.15, 9.50, 9.68, from redshift range 0.01$<$z$<$0.02 to 0.04$<$z$<$0.05. This
trend could be understand because both the $\alpha$.40 and SDSS MPA-JHU are the flux-limited samples, rather than volume-limited ones, 
thus only the brighter galaxies could be seen when the redshift raised. In each panel of this figure, we can see that either 
the star-forming \& composite galaxies (blue and green contours) or the AGNs (orange and red circles) display similar distributions 
as in Figure~\ref{fig2}, which indicate again that the dependency of M$_{HI}$, M$_{*}$ \& M$_{HI}$/M$_{*}$ on the AGN activity are negligible.

\subsection{IR Color \& AGN Activity}

Every method for searching active nuclei hidden in the center of galaxies, would inevitably lead to the bias to some particular 
types of AGNs. For instance, without considering the AGNs with obvious broad emission lines, the optical spectral diagnostic 
is biased to weak AGNs with powerless X-ray emission \citep{juneau13}. Moreover, only based on the selection of standard BPT 
diagrams, some early-type galaxies without apparent activity of nuclei but with prominent emissions from (post-) Asymptotic 
Giant Branch (AGB \& p-AGB) stars, would be possibly misdiagnosed as AGN-hosting galaxies \citep{stasinska08}. Hence, it is 
indispensable to adopt many methods to select AGNs and to quantify their activity. In fact, besides the optical emission lines, 
the detections of radio and hard X-ray continuum are common methods to search AGNs as well. Nevertheless, the radio continuum 
selection biases to radio-loud AGNs. Radio-loud AGNs are strong ones and the number of them is small with the fraction of 
about 10$\sim$20\% \citep[e.g.,][]{kellermann89,urry95}. 

Even though the hard X-ray luminosity is a powerful tool to identify nuclear accretion activity due to the ablility of hard 
X-ray photons to penetrate the obscuring material, this method is limited by the sensitivity of nowadays hard X-ray satellite 
\citep[e.g.,][]{tueller08}. The incremental Second XMM-Newton Serendipitous Source Catalog (2XMMi) is an updated version of 
the 2XMM catalog \citep{watson09}, and is the largest catalog of hard X-ray sources ever published so for. By matching 2XMMi 
and SDSS-DR7 photometric observations, \citet{pineau11} presented a catalog, which was used to join with the $\alpha$.40-SDSS 
sample in this study with the same SDSS PHOTOOBJID. Unfortunately, only 18 sources left, and only 1 of them is AGN (a Seyfert). 
Consequently, it is an unrealistic method to adopt the hard X-ray luminosity to trace the AGN activity for us. 

The output of galaxies at Infrared (IR) could be used to select AGNs as well. Besides IR spectral diagnostic \citep[e.g.,][]{genzel98, armus06, spoon07} 
and the IR bump at wavelength range from 10$\mu$m to 50$\mu$m \citep[e.g.,][]{laurent00,hao05,buchanan06,zhu08},
the slope of IR continuum at wavelength range from 1$\mu$m to 5$\mu$m is also sensitive to the activity of AGNs: generally, 
the continuum of a normal galaxy is dominated by evolved stellar populations at this wavelength range, and has a peak at approximately 
1.6$\mu$m, which is known as the 1.6$\mu$m bump feature \citep{hanami12}; while if a galaxy hosts a powerful AGN, the power-law 
spectral energy distributions (SEDs) would be expected as the most significant feature. Moreover, the extinction in this wavelength 
range is clearly weak. 
 
The methods by using the IR color criteria at wavelength of 1$\mu$m$<$$\lambda$$<$5$\mu$m to search AGNs, have been initiated 
by some studies \citep{lacy04, stern05} based on the observations of $Spitzer~Space~Telescope$ \citep{werner04}. Then, these 
methods have been developed \citep{stern12, assef13} on the basis of the observations of the Wide-field Infrared Survey Explorer 
\citep[WISE;][]{wright10}. Here, the IR color derived from WISE All-Sky Survey catalog will be utilized to check the optical 
spectral classification and quantify AGN activity.

WISE has observed the whole sky at $3.4$, $4.6$, $12$ and $22$$\mu$m with angular resolutions of 6.1$\arcmin$, 6.4$\arcmin$, 
6.5$\arcmin$ and 12.0$\arcsec$, respectively. 
precision has been improved by cross-matching with the catalog of 2MASS (stands for Two Micron All-Sky Survey). After doing 
that, the astrometric precision is better than 0.15 $\arcsec$ for the sources with higher S/N \citep{jarrett11}. The 6762 
galaxies in $\alpha$.40-SDSS sample were matched with those sources in WISE All-Sky Survey catalog with the radius of 3$\arcsec$, 
and 6270 of them have been observed by WISE with the S/N$>$3$\sigma$ at $3.4$ \& $4.6$$\mu$m. After the spectral classification 
described above, in this $\alpha$.40-SDSS-WISE sample, there are 4769 star-forming galaxies, 469 composite galaxies, 78 Seyferts 
(68 of them have EW(H$\alpha$)$<$-3.0\AA{}), 57 LINERs (18 of them have EW(H$\alpha$)$<$-3.0\AA{}). 

\begin{figure}
\figurenum{4}
\epsscale{1}
\plotone{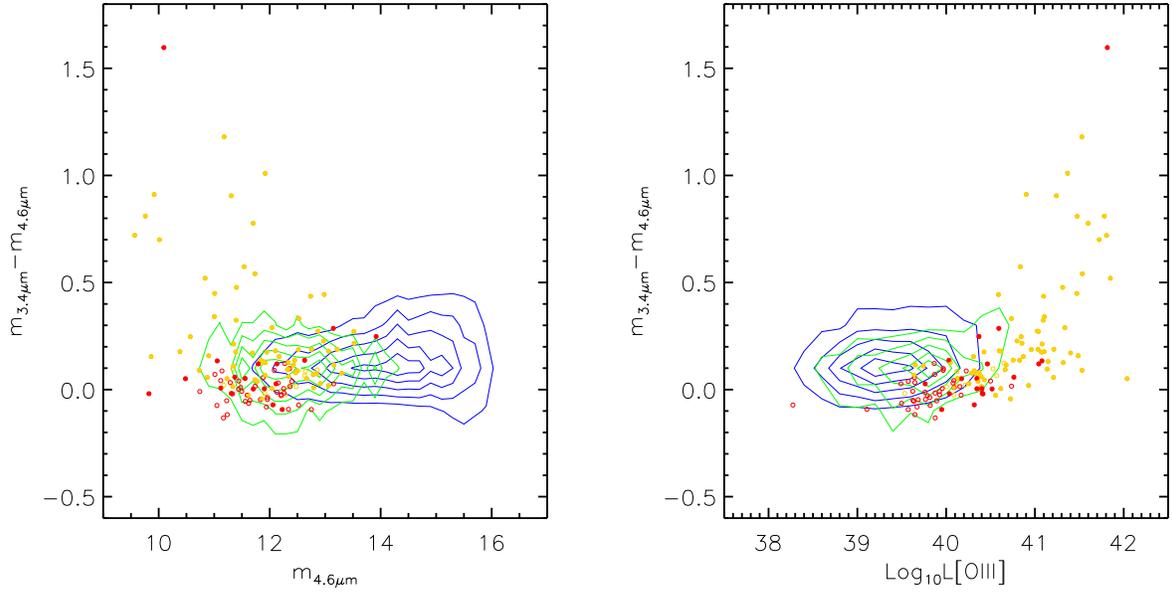}
\caption{Left panel is the color-magnitude diagram based on the observations of WISE. $m_{3.4 \mu m}$ \& $m_{4.6 \mu m}$ refer to the $3.4$ \& $4.6$~$\mu$m 
apparent magnitudes in Vega system. Right panel present the correlations between IR color and L$_{[OIII]}$. 
The denotations of the symbols and contours are the same as in Figure~\ref{fig2}. 
}
\label{fig4}
\end{figure}

The left panel of Figure~\ref{fig4} is a color-magnitude diagram to distinguish powerful AGNs from the galaxies whose continuums 
at wavelength of 1$\mu$m$<$$\lambda$$<$5$\mu$m are dominated by evolved stellar populations. The WISE magnitudes 
used here refer to the Vega system. \citet{assef13} have presented a criterion of $m_{3.4 \mu m}$-$m_{4.6 \mu m}$ $\geq$0.8 and W2$<$15.05 
to identify AGNs. In the left panel of Figure~\ref{fig4}, we could see that only a handful of optical selected AGNs could 
be classified as IR selected AGNs. Most of the optical selected AGNs, especially the LINERs, their outputs at $3.4$ \& $4.6$$\mu$m 
are definitely dominated by evolved stars, which cause their $m_{3.4 \mu m}$-$m_{4.6 \mu m}$ colors are approximately 0. The right panel of Figure~\ref{fig4} 
shows that the most of the sources with redder $m_{3.4 \mu m}$-$m_{4.6 \mu m}$ color, also have larger L$_{[OIII]}$.

\begin{figure}
\figurenum{5}
\epsscale{0.5}
\plotone{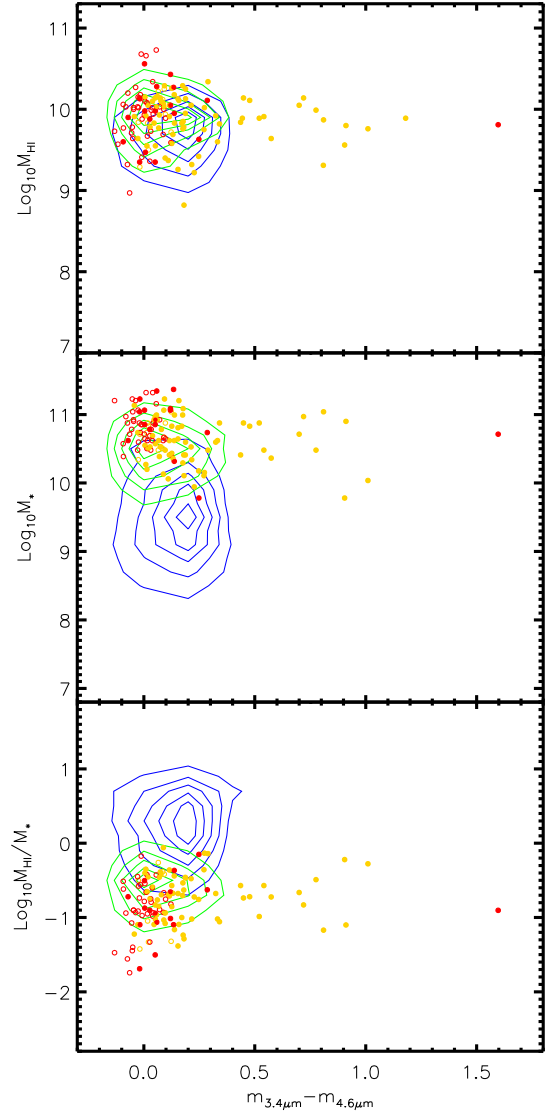}
\caption{Comparing the M$_{HI}$, M$_{*}$ \& M$_{HI}$/M$_{*}$ with IR color. 
The denotations of the symbols and contours are the same as in Figure~\ref{fig2}.}
\label{fig5}
\end{figure}

Figure~\ref{fig5} shows the variations of M$_{HI}$, M$_{*}$ \& M$_{HI}$/M$_{*}$ with the IR slope.
Apart from one galaxy, almost all the LINERs are crowded in a small region in each panel. The Seyferts' M$_{HI}$, M$_{*}$ 
\& M$_{HI}$/M$_{*}$ do not vary obviously with IR slope. Even for the powerful AGNs with red $m_{3.4 \mu m}$-$m_{4.6 \mu m}$ color, their neutral 
gas is still abundant, which is similar as the result obtained above.



\subsection{AGN Effect on Neutral Gas}

The beginning of star formation is a key problem for understanding the evolution of galaxies. As the most abundance element 
in the universe, HI is the raw material of stars \citep{schmidt59}. Nevertheless, HI is not direct connect with star formation 
\citep{waller87, kennicutt98b}. This kind of gas still needs to cool down enough to condense to form stars \citep{mckee07, zinnecker07}. 
Observations have indicate that neutral atomic gas have transfered to molecular gas before collapsing and fragmenting to form 
stars \citep{gao04, brodie06, komugi07, bigiel08}. The quenching of the star formation is important as well. The most popular 
candidates to extinct the forming of stars are the environmental effects \citep{smith12, tal14} and the feedback from the supermassive black holes \citep{dimatteo05, hopkins05}.

In this work, we confirm that almost all the type-II AGNs are gas-rich, similar as the results presented by \citet{konig09, ho08b, fabello11}. 
And the AGN activity, traced by L$_{[OIII]}$, d$_{SF}$ and IR color, seems not to influence the reservoir of neutral gas. 
We should note that the AGN-hostings in $\alpha$.40-SDSS sample are mature galaxies. The young stellar populations are absent 
in these sources (see next sub-section). 
So the quenching must have happened million years ago, and AGN feedback would be responsible for the quenching.

Whatever, it is surprised that plentiful gas is still possessed by the quenched galaxies. Without considering the environmental 
effects, though they can not be ruled out roughly, a plausible explanation is the power of AGN feedback is too small to expel the gas, which has been introduced by \citet{ho08b}. They speculated the similarity of the HI abundance between AGN and 
the inactive galaxies, may be due to the fact that the feedback of black holes could only expel gas close to the centers of 
the galaxies, but can not heat the gas that spread all over the galaxies. On the basis of this hypothesis, the AGN feedback 
should be confined to affecting the gas surrounding the jets or winds with narrow solid angle. The AGN-related gaseous outflow 
might also be detected possibly, but the total amount of outflow should be insignificant. If this assumption is really true, 
it is would be a challenge to interpret the simultaneous suppression of star formations, because the distribution of star 
formation regions is not barely restricted in the central part of galaxies except those with violent starburst \citep{kennicutt12}.

The puzzle (the AGNs lodge in gas-rich galaxies) could also be solved by some other possible explanations. 
Firstly, if the AGN feedback does not push the gas out of the galaxies, but just heats the gas too warm and sparse to condense 
the embryos of stars; or, if the feedback just prevents the continuant inflow of fresh intergalactic gas, which is generally 
considered as an important process for the disk galaxies to form stars \citep{chiappini97, chiappini01}, both these two assumptions 
would lead to the observational fact that AGNs are hosted in the gas-rich mature galaxies without on-going star-forming. 
Alternatively, we should notice that the beam of ALFALFA survey is about 3.5$\arcmin$, corresponding to a range of 40 to 200kpc 
from redshift at z$\sim$0.01 to 0.05, which is a much larger coverage than the visible size of normal galaxies. Hence, if 
the neutral gas content has been pushed out of the visible boundary of the hosts by the AGN feedback, yet not out of the gravitational potential well controled by the dark matter halos, we could not point out where the gas is. In this case, we 
can not ascertain the power of AGN feedback as well. Provided that the cold gas has really been pushed out to the suburbs of 
the AGN-hostings, the star formation may be still exist on the periphery. This supposition is in accordance with the hypothesis 
of inside-out formation of galactic disks \citep{fraternali12}, but can not been confirmed by spectral observations of SDSS 
due to the 3$\arcsec$ diameter fiber. Actually, based on the observations of integral field unit (IFU), \citet{pracy14} have 
found ongoing star formation outside the center of two nearby E+A galaxies, while their central regions are absent of ionizing stars. 

The coexistence of massive cold gas and quenching from AGN feedback (or environment effects, which can not be ruled out but 
do not be considered in this work) in one galaxy is a challenge for current hypotheses about galaxy evolution. All the assumptions 
listed above can lead to the fact that the gas-rich mature galaxies host an active nucleus. But on the basis of available 
HI observations, it is very hard to ascertain which assumption is the principle one. The influence of AGN feedback on the surrounding gas has been smoothed down due to the too large coverage of nowaday HI survey. Eventually, if one of the assumptions 
is proved to be the truth, it is very possible that we have underestimated the power of AGN feedback. In order to solve this 
problem, deep HI imagings with much better angular resolution of diverse galaxies in large local samples are needful in the 
future, which could be used to detect the variation of radial surface brightness of neutral gas. Alternatively, since the 
molecular gas is denser and more concentrated than neutral gas \citep{bigiel08}, the connections between AGN activity and 
molecular gas in the central region of galaxies could be investigated detailedly as well.

\subsection{The Bias of HI Selected Galaxies}

The limitations are exist for HI surveys in terms of HI line emission, and some studies have discussed this kind of bias \citep{west10,martin12}. 
On the basis of the $\alpha$.40 data and by using the color-magnitude diagram \citep{baldry04} to classify them with optical 
color, \citet{huang12} found the major HI selected galaxies are the ones belong to the blue cloud, while the optical red galaxies 
are rare in their sample. This means the selection of HI observations are biased to the galaxies with current star formation. 
As a by-product, the bias of blind HI survey on the spectral classifications could been presented in this work.

For the various types of galaxies shown in Table 1, the fractions detected by ALFALFA are significant different. 
As expected, the star-forming galaxies are the ones discovered most easily by HI survey, with a detection percentage of 37.7\%. 
For the other four types of galaxies, composites, AGNs (without considering the S/N of [OI] here), transitions \& absorptions, 
the fractions detected by ALFALFA are 32.3\%, 29.7\%, 18.6\% \& 2.1\%, respectively. For the Seyferts and LINERs with different 
EW(H$\alpha$) criteria, there is no valuable variation in their HI detections (see Table 2). In a word, HI surveys are sensitive 
to the galaxies with significant optical emission lines, and the invisibility for HI observations is only confined to the 
galaxies with absorption or extremely weak emission lines. 

\begin{figure}
\figurenum{6}
\epsscale{1}
\plotone{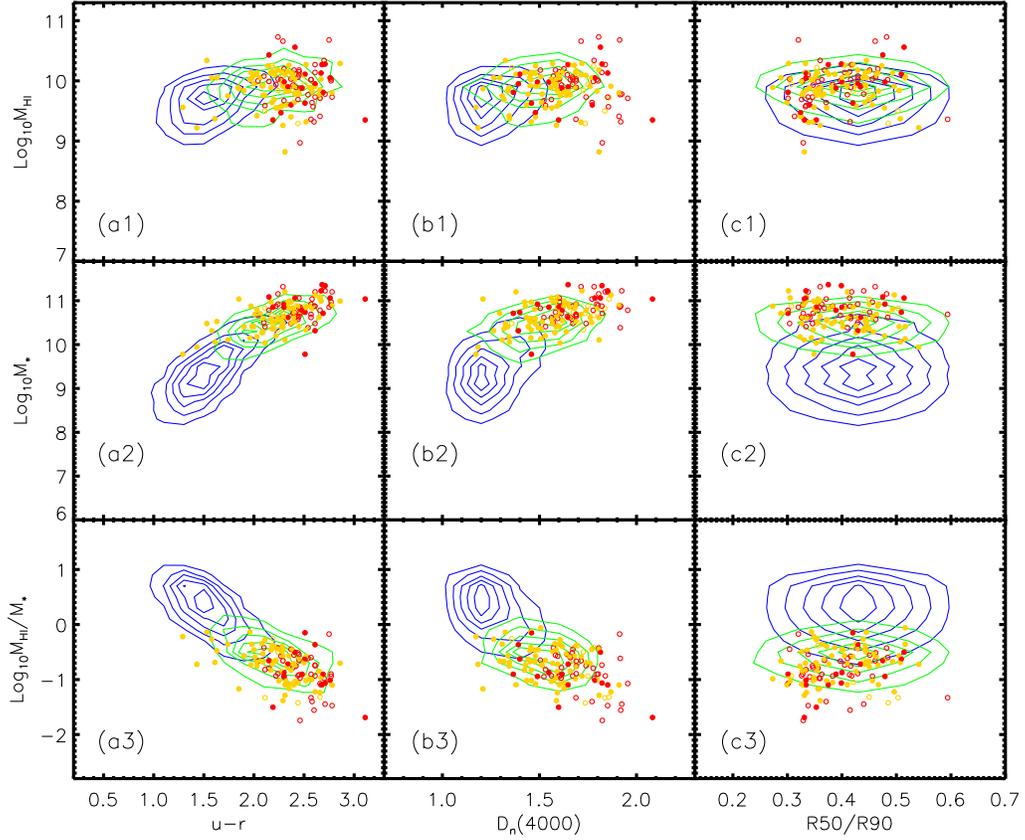}
\caption{Comparing the M$_{HI}$, M$_{*}$ \& M$_{HI}$/M$_{*}$ with various factors, including u-r color (a1, a2, 
a3), $D_n$(4000) (b1, b2, b3), $R50/R90$ (c1, c2, c3). The denotations of the symbols and contours are the same as in Figure~\ref{fig2}.
}
\label{fig6}
\end{figure}

The absorption line and weak emission line galaxies are generally old galaxies with red color, in contrast with star-formings 
whose optcial appearances are usually blue, which means they have different star formation history (SFH). The hosts of weak 
AGNs are usually considered as transitional or interim galaxies between old red and young blue galaxies \citep{coil08, zehavi11, mendez11}. 
Therefore, it would be beneficial to study the correlations between HI content and SFH in different types of galaxies. Figure~\ref{fig6} 
shows the M$_{HI}$, M$_{*}$ \& M$_{HI}$/M$_{*}$ as a function of various factors, including $u$-$r$ color (a1, a2, a3), $D_n$(4000) 
(b1, b2, b3), $R50/R90$ (c1, c2, c3). Here, the model magnitudes at $u$ \& $r$ bands are used, and the K-correction that has 
been performed was based on the IDL code supplied by Blanton (version tag v4\_1\_4). The method and the SEDs used by this code 
were described by \citet{blanton03} and \citet{blanton07}. The 4000\AA{} break index $D_n$(4000) is widely used as an excellent 
age indicator of the stellar population of a galaxy until a few gigayears after the onset of star formation activity \citep{bruzual03, kauffmann03b, kauffmann03c}. 
Here, both of $u$-$r$ color and $D_n$(4000) represent the SFH of galaxies. The $r$ band concentration parameters $R$50/$R$90 
is defined as ratio of two radii containing 50\% and 90\% of the Petrosian $r$ band luminosity, and is often used to represent 
the morphology of galaxies \citep{kauffmann03b, kauffmann03c}.

\begin{figure}
\figurenum{7}
\epsscale{1}
\plotone{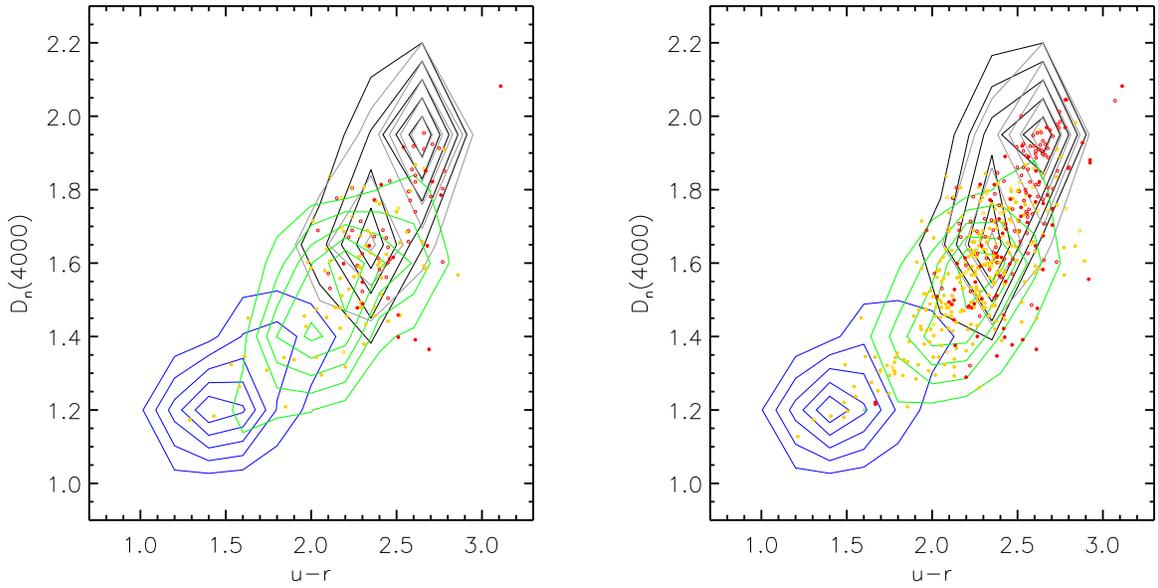}
\caption{The correlations between u-r color and $D_n$(4000) for the galaxies belong to $\alpha$.40-SDSS sample (left) or SDSS 
sample (right). The solid orange \& red circles in both panels represent the Seyferts and LINERs with EW(H$\alpha$)$<$-3.0\AA{}; 
the open orange \& red circles represent the Seyferts and LINERs with -3.0\AA$<$EW(H$\alpha$)$<$-1.0\AA; 
Blue and Green contours denote the locations of star-forming galaxies and composites; 
while the gray and black contours denote the locations of transitions and absorptions, respectively. }
\label{fig7}
\end{figure}

In Figure~\ref{fig6}, the morphology of galaxies, represented by $R$50/$R$90, do not show correlations with the M$_{HI}$, 
M$_{*}$ \& M$_{HI}$/M$_{*}$. M$_{HI}$ changes slowly with $u$-$r$ color \& $D_n$(4000), but M$_{*}$ and M$_{HI}$/M$_{*}$ 
are related with SFH closely: younger galaxies have less M$_{*}$ and larger M$_{HI}$/M$_{*}$, and vice versa. We can see that 
the AGN-hostings have red color and large $D_n$(4000), thus they are definitely older than star-forming galaxies. 
In order to verify this conclusion again, the correlations between $u$-$r$ color \& $D_n$(4000) are shown in Figure~\ref{fig7}. 
Besides star-formings \& composites \& AGNs, the transition and absorption line galaxies are also demonstrated in Figure~\ref{fig7} 
with gray and black contours. The left and right panel of Figure~\ref{fig7} show the galaxies in the $\alpha$.40-SDSS \& SDSS sample, respectively. The SDSS sample contains more transition and absorption line galaxies, while these two types of 
galaxies are rare in the $\alpha$.40-SDSS sample. It is easy to find that, apart from some galaxies (all of them are Seyferts), 
most of the AGNs have the similar $u$-$r$ color \& $D_n$(4000) with transition and absorption line galaxies. 

AGN-hostings have significant higher HI detection than transition and absorption line galaxies, but they three types of galaxies 
show similar SFH. This means, for the galaxies with similar stellar populations, AGN-hostings maintain more HI content than 
the ones without AGN. If the central engine of an AGN-hosting has extinguished, only a red galaxy possessed by evolved stars 
could be observed. However, as discussed above, the AGN-hostings are gas-rich and the AGNs have not sufficient power to expel 
gas content in their hosts. Thus, if an AGN-hosting would ultimately migrate to an old red galaxy with few cold gas, such 
as the absorption line galaxy in this work, some questions that emerge are: when and how does the AGN-hosting exhaust its 
gas content? Does the AGN quenching correlate with evolution stage rather than AGN activity? Are there star formation regions  
on the periphery of AGN-hostings, and how much of cold gas could be consumed by them? Besides AGN, are there other quenching 
processes that could be responsible for the exhausting of gas? The detailed investigations to these questions are definitely 
beyond the scope of this work. But solving these problems are essential for understanding the evolution of galaxies. 

\subsection{LINERs}

The similarity of LINERs and the galaxies dominated by old stellar populations has been mentioned in previous sub-section.
As a kind of dominant AGN (if it is) in local universe \citep{ho97}, the properties of LINERs should be studied detailedly.
The physical origin of LINERs has long been the subject of hot debate ever since they were identified by \citet{heckman80}. 
For this type of galaxies, with relatively weak emission line, can be produced by a wide array of ionization mechanisms \citep{wang09,annibali10}. 
One of the origins for LINERs that have been proposed is low accretion-rate AGN \citep{ferland83,halpern83}. The galaxies 
with hot evolved stars, such as AGB \& p-AGB, might also contribute to the ionization budget \citep{terlevich85,binette94,taniguchi00}. 
The fast shocks driven by supernovae or stellar eject, is another major explanation to LINERs \citep{heckman80, dopita95, dopita97}. 

The role of LINERs has been controversial, and one of the disputing points is the exist of central active nucleus or not. 
The explanation with low accretion-rate AGN is strongly supported by the detection of broad emission lines in the optical 
spectra in some of local LINERs \citep{ho97, ho08c}. Other evidences, on the basis of observations from hard X-ray to radio, 
also support this standpoint \citep{nagar05,gonzalez09,terashima00}. Nevertheless, recently, some inconsistencies have been 
found between observaions and the AGN-ionization hypothesis: \citet{capetti11a} \& \citet{capetti11b} found close connection 
between emission lines and stellar continuum; \citet{yan12} find the line emission in the majority of LINERs is spatially 
extended, which has also been supported by the observations with IFU \citep{bremer13,sarzi10,singh13}. 
Approximately all these studies proposed that the AGB \& p-AGB stars are the ionization sources \citep{trinchieri91}. 

\begin{figure}
\figurenum{8}
\epsscale{1}
\plotone{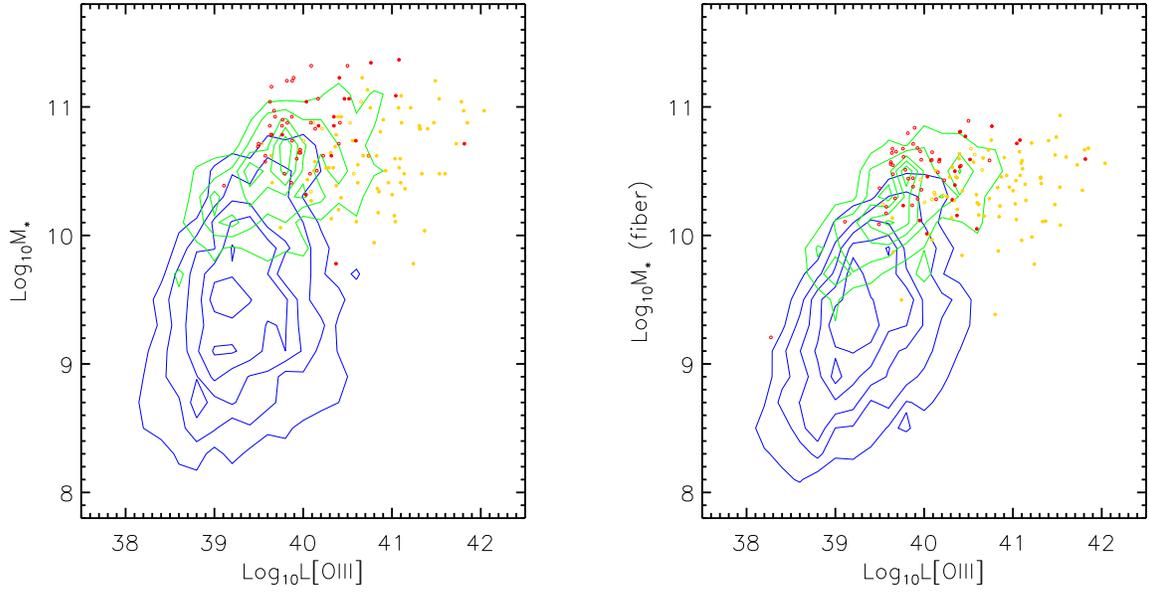}
\caption{
The correlations between stellar mass and L$_{[OIII]}$. The stellar mass used in the left panel is the total stellar mass (M$_{*}$) 
of the galaxies; while the stellar mass used in the right panel is the stellar mass in the area coverd by 3$\arcsec$ diameter 
fiber of SDSS (M$_{*}$(fiber)). The denotations of the symbols and contours are the same as in Figure~\ref{fig2}. 
}
\label{fig8}
\end{figure}

Compared with Seyferts, Figure~\ref{fig7} shows more similarity between LINERs and the galaxies dominated by old stellar populations.
In order to test it again, the correlations between L$_{[OIII]}$ and the stellar mass are presented in Figure~\ref{fig8}. 
In the left panel, the stellar mass, M$_{*}$, was derived from the MPA-JHU catalog to represent the total stellar contents of the galaxies. 
This panel is same as the panel (a2) of Figure~\ref{fig2}. In the right panel of Figure~\ref{fig8}, the stellar mass was estimated 
by using the fiber photometries of SDSS on the basis of the equation supplied by \citet{bell03}, who presented the relationship 
between various optical colors and mass-to-luminosity ratios. Therefore, the stellar mass (hereafter as M$_{*}$(fiber)) in the right panel of Figure~\ref{fig8} 
represents the stellar contents of the galaxies covered by the 3$\arcsec$ diameter fiber of SDSS. 
The representations of the symbols and contours are the same as in Figure~\ref{fig2}. In the right panel of Figure~\ref{fig8}, 
a correlation between M$_{*}$(fiber) and L$_{[OIII]}$ could be seen for the star-formings (blue contour) and composites (green 
contour). Approximately all the LINERs (red open \& solid circles) show the same correlation with star-formings and composites. 
But most of Seyferts (orange circles) show obvious higher L$_{[OIII]}$ than the others, and only a small amount of Seyferts, 
especially the ones with -3.0\AA$<$EW(H$\alpha$)$<$-1.0\AA{} (open orange circles), follow the correlation defined by 
star-formings and composites. Hence, this may be another evidence for LINERs hosting non-AGN ionization sources. However, 
it must be noted that almost all the evidences listed above, including the evidences in other studies, can not ruled out the 
possibility of a weak active nucleus in the central part of LINER; maybe the power of the AGN is barely confined to a significant small region.

\citet{fabello11} has found that the M$_{HI}$/M$_{*}$ is a function of the Eddington ratio for the optical selected AGNs 
(include Seyferts, LINERs and composites). The Eddington ratio they used are traced by L$_{[OIII]}$/$\sigma$$^{4}$, and $\sigma$ 
represent the stellar velocity dispersion. Nevertheless, the denominators of both L$_{[OIII]}$/$\sigma$$^{4}$ and M$_{HI}$/M$_{*}$ 
are dominated by old stellar populations. Hence, if the [OIII] emission is ionized by stars instead of AGNs, a similar 
correlation between L$_{[OIII]}$/$\sigma$$^{4}$ and M$_{HI}$/M$_{*}$, such as the one obtained by \citet{fabello11}, 
could be expected as well. In this context, the correlation between them do not reflect the influence of AGNs at all. 


\section{Summary}
Based on a sample of galaxies selected from $\alpha$.40 catalog, the dependence of the M$_{HI}$, M$_{*}$ \& M$_{HI}$/M$_{*}$ 
on various tracers of AGN activity are presented and analyzed. The main results described in this paper can be summarized as follows.

1. Approximately all the AGN-hosting galaxies (both type-II Seyferts and LINERs) in $\alpha$.40-SDSS sample are gas-rich galaxies. 
The neutral gas content in AGN-hostings is as abundant as in star-forming and composite galaxies. 
But the M$_{HI}$/M$_{*}$ of AGN-hostings is lower than star-formings' because the former contains more stars.

2. For the AGN-hosting galaxies (both type-II Seyferts and LINERs), there is no evidence to indicate the dependence of M$_{HI}$, 
M$_{*}$ \& M$_{HI}$/M$_{*}$ with the AGN activity: The M$_{HI}$ and M$_{*}$ in these sources do not increase or decrease obviously 
with the raising of L$_{[OIII]}$, d$_{SF}$ \& IR slope.

3. It is puzzled that the AGNs lodge in gas-rich galaxies. But the beam size of ALFALFA survey is about 3.5$\arcmin$, corresponding 
to 40 to 200kpc from redshift at z$\sim$0.01 to 0.05, definitely larger than the visible size of majority of galaxies. If 
the neutral gas content has been pushed out of the visible boundary of the hosts by the AGN feedback, yet not out of the gravitational 
potential well controled by the dark matter halos, we could not point out where the gas is.
More HI observations with much better angular resolutions are indispensable to solve this puzzle. 

4. HI surveys are sensitive to the galaxies with significant optical emission lines, and the invisibility for HI observations 
is only confined to the galaxies with absorption or extremely weak emission lines. AGN-hostings shown similar SFH with transition 
and absorption line galaxies, but the former have significant higher HI detection. Therefore, if an AGN-hostings would ultimately 
migrate to an old red galaxy with few cold gas, some puzzles, such as when and how the AGN-hosting exhaust its gas content, 
should be solved by future hypotheses and observaitons.

5. LINERs in $\alpha$.40-SDSS sample prefer to contain more stars than Seyferts, and LINERs follow the correlations between 
L$_{[OIII]}$ and stellar mass defined by star forming and composite galaxies. These may be another evidence for LINERs hosting 
non-AGN ionization sources.

\acknowledgments
The author acknowledges Dr. Z.-M. Zhou and M.-I. Lam for advice and helpful discussions. 
This project is supported by the Ministry of Science and Technology of the People's republic of China through National Basic Research Program of China (973 program) No. 2012CB821800, the National Natural Science Foundation of China through grant 11173030, 11225316, 11078017 and 10833006.

The Arecibo Observatory is part of the National Astronomy and Ionosphere Center, which is operated by Cornell University under a cooperative agreement with the National Science Foundation.The authors acknowledge the work of the entire ALFALFA collaboration team in observing, flagging, and extracting the catalog of galaxies used in this work. The ALFALFA team at Cornell is supported by NSF grant AST-0607007 and AST-1107390 and by grants from the Brinson Foundation.
 
The author thank the useful SDSS database and the MPA/JHU catalogs.
Funding for the SDSS and SDSS-II has been provided by the Alfred P. Sloan Foundation, the Participating Institutions,
the National Science Foundation, the U.S. Department of Energy, the National Aeronautics and Space Administration, the Japanese
Monbukagakusho, the Max Planck Society, and the Higher Education Funding Council for England. The SDSS Web Site is http://www.sdss.org/.
The SDSS is managed by the Astrophysical Research Consortium for the Participating Institutions. The Participating Institutions are the
American Museum of Natural History, Astrophysical Institute Potsdam, University of Basel, University of Cambridge, Case Western
Reserve University, University of Chicago, Drexel University, Fermilab, the Institute for Advanced Study, the Japan Participation
Group, Johns Hopkins University, the Joint Institute for Nuclear Astrophysics, the Kavli Institute for Particle Astrophysics and
Cosmology, the Korean Scientist Group, the Chinese Academy of Sciences (LAMOST), Los Alamos National Laboratory, the Max-Planck-Institute
for Astronomy (MPIA), the Max-Planck-Institute for Astrophysics (MPA), New Mexico State University, Ohio State University, University
of Pittsburgh, University of Portsmouth, Princeton University, the United States Naval Observatory, and the University of Washington.

This publication makes use of data products from the Wide-field Infrared Survey Explorer, which is a joint project of the 
University of California, Los Angeles and the Jet Propulsion Laboratory/California Institute of Technology, funded by the 
National Aeronautics and Space Administration.

This work is also based on observations obtained with XMM-Newton, an ESA science mission with instruments and contributions directly funded by ESA Member States and the USA (NASA).

\clearpage


\clearpage

\clearpage
 
\clearpage

\clearpage

\clearpage
 
\clearpage

\clearpage

\clearpage

\clearpage


\begin{deluxetable}{lcccccc}
\centering
\tablecolumns{7}
\tabletypesize{\footnotesize}
\tablewidth{0pt}
\tablecaption{The numbers of samples with different spectral types}
\tablehead{
\colhead{Sample} & \colhead{$star-forming$} & \colhead{$composite$} & 
\colhead{$AGN$} & \colhead{$AGN([OI])$} & \colhead{$transition$} & \colhead{$absorption$}}
\startdata
$\alpha$.40-SDSS   & 5162 & 490 & 153 & 141 &  485 &   72 \\
    SDSS           &13706 &1517 & 515 & 460 & 2608 & 3389 \\
     \%            & 37.7 & 32.3& 29.7& 30.6& 18.6 &  2.1 \\ 

\hline
\enddata
\tablecomments{
The third row is the fraction numbers detected by ALFALFA survey. Col.(4): AGN number based on selection of [NII]/H$\alpha$ versus [OIII]/H$\beta$ diagram; Col.(5): AGN number with the S/N of the flux of [OI] emission lines $>$3$\sigma$.
}

\label{tab1}
\end{deluxetable}

\begin{deluxetable}{lcccc}
\centering
\tablecolumns{5}
\tabletypesize{\footnotesize}
\tablewidth{0pt}
\tablecaption{The numbers of Seyferts \& LINERs with different EW(H$\alpha$)}
\tablehead{
\colhead{Sample} & \colhead{Seyf ($<$-1.0\AA{})} & \colhead{Seyf ($<$-3.0\AA{})} &
\colhead{LINER ($<$-1.0\AA{})} & \colhead{LINER ($<$-3.0\AA{})}}

\startdata
$\alpha$.40-SDSS   & 80 & 70 &  58 &  18  \\
 SDSS              &273 &226 & 174 &  69  \\
     \%            &29.3&31.0& 33.3& 26.1 \\ 
\hline
\enddata
\tablecomments{    
Same as Table 1, the third row is the fraction numbers detected by ALFALFA survey.
Col.(2) \& (4): the numbers of Seyferts \& LINERs with EW(H$\alpha$)$<$-1.0\AA{}, the minus sign means emission; 
Col.(3) \& (5): the numbers of Seyferts \& LINERs with EW(H$\alpha$)$<$-3.0\AA{}.
}

\label{tab2}
\end{deluxetable}


\begin{thebibliography}{}
\bibitem[Abazajian et al.(2009)]{abazajian09} Abazajian, K. N., Adelman-McCarthy, J. K., Ag{\"u}eros, M. A., Allam, S. S., et al.\ 2009, \apjs, 182, 543
\bibitem[Annibali et al.(2010)]{annibali10} Annibali, F., Bressan, A., Rampazzo, R., Zeilinger, W. W., et al.\ 2010, \aap, 519, 40
\bibitem[Antonucci (1993)]{antonucci93} Antonucci R.\ 1993, \araa, 31, 473
\bibitem[Armus et al.(2006)]{armus06} Armus, L., Bernard-Salas, J., Spoon, H. W. W., Marshall, J. A, et al.\ 2006, \apj, 640, 204
\bibitem[Assef et al.(2013)]{assef13} Assef, R. J., Stern, D., Kochanek, C. S., Blain, A. W. et al.\ 2013, \apj, 772, 26
\bibitem[Baldwin et al.(1981)]{baldwin81} Baldwin, J. A., Phillips, M. M., \& Terlevich, R.\ 1981, \pasp, 93, 5B
\bibitem[Baldry et al.(2004)]{baldry04} Baldry, I. K., Glazebrook, K., Brinkmann, J., Ivezi{\'c}, \u{Z}., et al.\ 2004, \apj, 600, 681
\bibitem[Barnes et al.(2001)]{barnes01} Barnes, D. G., Staveley-Smith, L., de Blok, W. J. G., Oosterloo, T., et al.\ 2001, \mnras, 322, 486
\bibitem[Bell et al.(2003)]{bell03} Bell, E. F., McIntosh, D. H., Katz, N., \& Weinberg, D.\ 2003, \apjs, 149, 289
\bibitem[Bergin \& Tafalla (2007)]{bergin07} Bergin, E. A. \& Tafalla, M.\ 2007, \araa, 45, 339
\bibitem[Bigiel et al.(2008)]{bigiel08} Bigiel, F., Leroy, A., Walter, F., Brinks, E., et al.\ 2008, \aj, 136, 2846
\bibitem[Binette et al.(1994)]{binette94} Binette, L., Magris, C. G., Stasi{\'n}ska, G.; Bruzual, A. G.\ 1994, \aap, 292, 13
\bibitem[Blanton et al.(2003)]{blanton03} Blanton, M. R., et al.\ 2003, \aj, 125, 2348
\bibitem[Blanton \& Roweis (2007)]{blanton07} Blanton, M. R. \& Roweis, S.\ 2007, \aj, 133, 734
\bibitem[Boselli et al.(2010)]{boselli10} Boselli, A., Eales, S., Cortese, L., Bendo, G., et al.\ 2010, \pasp, 122, 261
\bibitem[Bremer et al.(2013)]{bremer13} Bremer, M., Scharw{\"a}chter, J., Eckart, A., Valencia-S., M.\ 2013, \aap, 558, 34
\bibitem[Brodie \& Strader (2006)]{brodie06} Brodie, J. P. \& Strader, J.\ 2006, \araa, 44, 193
\bibitem[Bruzual \& Charlot (2003)]{bruzual03} Bruzual, G. \& Charlot, S.\ 2003, \mnras, 344, 100
\bibitem[Buchanan et al.(2006)]{buchanan06} Buchanan, C. L., Gallimore, J. F., O'Dea, C. P., Baum, S. A., et al.\ 2006, \aj, 132, 401
\bibitem[Cardelli et al.(1989)]{cardelli89} Cardelli, J. A., Clayton, G. C., \& Mathis, J. S.\ 1989, \apj, 345, 245
\bibitem[Calzetti (2001)]{calzetti01} Calzetti, D.\ 2001, \pasp, 113, 1449
\bibitem[Capetti \& Baldi (2011)]{capetti11a} Capetti, A. \& Baldi, R. D.\ 2011, \aap, 529, 126
\bibitem[Capetti (2011)]{capetti11b} Capetti, A. \ 2011, \aap, 535, 28
\bibitem[Chiappini et al.(1997)]{chiappini97} Chiappini, C., Matteucci, F. \& Gratton, R.\ 1997, \apj, 477, 765
\bibitem[Chiappini et al.(2001)]{chiappini01} Chiappini, C., Matteucci, F. \& Romano, D.\ 2001, \apj, 554, 1044
\bibitem[Cid Fernandes et al.(2010)]{cid10} Cid Fernandes R., Stasi{\'n}ska, G., Schlickmann, M. S., Mateus, A., et al.\ 2010, \mnras, 403, 1036
\bibitem[Cid Fernandes et al.(2011)]{cid11} Cid Fernandes R., Stasi{\'n}ska, G., Mateus, A. \& Vale Asari, N.\ 2011, \mnras, 413, 1687
\bibitem[Coil et al.(2008)]{coil08} Coil, A. L., Newman, J. A., Croton, D., Cooper, M. C., et al.\ 2008, \apj, 672, 153
\bibitem[Cowie et al.(2003)]{cowie03} Cowie, L. L., Barger, A. J., Bautz, M. W., Brandt, W. N., \& Garmire, G. P.\ 2003, \apj, 584, 57
\bibitem[Di Matteo et al.(2005)]{dimatteo05} Di Matteo T., Springel V. \& Hernquist L.\ 2005, Nature, 433, 604
\bibitem[Dopita \& Sutherland (1995)]{dopita95} Dopita, M. A. \& Sutherland, R. S.\ 1995, \apj, 455, 468
\bibitem[Dopita et al.(1997)]{dopita97} Dopita, M. A., Koratkar, A. P., Allen, M. G., Tsvetanov, Z. I., et al.\ 1997, \apj, 490, 202
\bibitem[Evoli et al.(2011)]{evoli11} Evoli, C., Salucci, P., Lapi, A. \& Danese, L., et al.\ 2011, \apj, 743, 45
\bibitem[Fabello et al.(2011)]{fabello11} Fabello, S., Kauffmann, G., Catinella, B., Giovanelli, R, et al.\ 2011, \mnras, 416, 1739
\bibitem[Fabian (1999)]{fabian99} Fabian, A. C.\ 1999, \mnras, 308, 39
\bibitem[Farage et al.(2010)]{farage10} Farage, C. L., McGregor, P. J., Dopita, M. A., \& Bicknell, G. V.\ 2010, \apj, 724, 267
\bibitem[Ferland \& Netzer (1983)]{ferland83} Ferland, G. J. \& Netzer, H.\ 1983, \apj, 264, 105
\bibitem[Fraternali \& Tomassetti (2012)]{fraternali12}Fraternali, F. \& Tomassetti, M.\ 2012, \mnras, 426, 2166
\bibitem[Gao \& Solomon (2004)]{gao04} Gao, Y. \& Solomon, P. M.\ 2004, \apj, 606, 271
\bibitem[Gonz{\'a}lez-Mart{\'i}n et al.(2009)]{gonzalez09} Gonz{\'a}lez-Mart{\'i}n, O., Masegosa, J., M{\'a}rquez, I., \& Guainazzi, M.\ 2009, \apj, 704, 1570
\bibitem[Genzel et al.(1998)]{genzel98} Genzel, R., Lutz, D., Sturm, E., Egami, E., et al.\ 1998, \apj, 498, 579
\bibitem[Giovanelli et al.(2005a)]{giovanelli05a} Giovanelli, R., Haynes, M. P., Kent, B. R., Perillat, P., et al.\ 2005, \aj, 130, 2613
\bibitem[Giovanelli et al.(2005b)]{giovanelli05b} Giovanelli, R., Haynes, M. P., Kent, B. R., Perillat, P., et al.\ 2005, \aj, 130, 2598
\bibitem[Giovanelli et al.(2007)]{giovanelli07} Giovanelli, R., Haynes, M. P., Kent, B. R., Saintonge, A., et al.\ 2007, \aj, 133, 2569
\bibitem[Granato et al.(2004)]{granato04} Granato, J. E., et al.\ 2004, \apj, 600, 580
\bibitem[Greene \& Ho (2006)]{greene06} Greene, J. E. \& Ho, L. C.\ 2006, \apj, 641, 21
\bibitem[Halpern \& Steiner (1983)]{halpern83}Halpern, J. P. \& Steiner, J. E. 1983, \apj, 269, L37
\bibitem[Hanami et al.(2012)]{hanami12} Hanami, H., Ishigaki, T., Fujishiro, N., Nakanishi, K,  et al.\ 2012, \pasj, 64, 70
\bibitem[Hao et al.(2005)]{hao05} Hao, C.-N., Xia, X.-Y., Mao, S., Wu, H. \& Deng, Z.-G.\ 2005, \apj, 625, 78
\bibitem[Hatziminaoglou et al.(2009)]{hatziminaoglou09} Hatziminaoglou, E., Fritz, J. \& Jarrett, T. H.\ 2009, \mnras, 399, 1206
\bibitem[Haynes et al.(2007)]{haynes07} Haynes, M. P., Giovanelli, R., \& Kent, B. R.\ 2007, \apj, 665, 19
\bibitem[Haynes et al.(2011)]{haynes11} Haynes, M. P., Giovanelli, R., Martin, A. M., \& Hess, K. M., et al.\ 2011, \aj, 142, 170
\bibitem[Heckman et al.(1978)]{heckman78} Heckman, T. M., Balick, B. \& Sullivan, W. T., III \ 1978, \apj, 224, 745
\bibitem[Heckman (1980)]{heckman80} Heckman, T. M.\ 1980, \aap, 87, 152
\bibitem[Heckman et al.(2004)]{heckman04} Heckman T. M., Kauffmann G., Brinchmann J., Charlot S., Tremonti C., \& White S. D. M.\ 2004, \apj, 613, 109
\bibitem[Ho et al.(1997)]{ho97} Ho, L. C., Fillipenko, A. V., \& Sargent, W. L. W.\ 1997, \apj, 487, 568
\bibitem[Ho et al.(2008a)]{ho08a} Ho, L. C., Darling, J., \& Greene, J. E.\ 2008, \apjs, 177, 103
\bibitem[Ho et al.(2008b)]{ho08b} Ho, L. C., Darling, J., \& Greene, J. E.\ 2008, \apj, 681, 128
\bibitem[Ho (2008)]{ho08c} Ho, L. C.\ 2008, \araa, 46, 475
\bibitem[Hogg et al.(2005)]{hogg05} Hogg, D. W., Tremonti, C. A., Blanton, M. R., Finkbeiner, D. P., et al.\ 2005, \apj, 624, 162
\bibitem[Hopkins A. et al.(2003)]{hopkins03} Hopkins, A. M., Miller, C. J., Nichol, R. C., Connolly, A. J., et al.\ 2003, \apj, 599, 971
\bibitem[Hopkins A. et al.(2005)]{hopkins05} Hopkins, P. F., Hernquist, L., Martini, P., Cox, T. J., et al.\ 2005, \apj, 625, 71 
\bibitem[Hopkins P. et al.(2006)]{hopkins06} Hopkins, P. F., Hernquist, L., Cox, T. J., Di Matteo, T., et al.\ 2006, \apjs, 163, 1
\bibitem[Huang et al.(2012)]{huang12} Huang, S., Haynes, M. P., Giovanelli, R. \& Brinchmann, J.\ 2012, \apj, 756, 113
\bibitem[Hughes \& Cortese (2009)]{hughes09} Hughes, T. M. \& Cortese, L.\ 2009, \mnras, 396, 41
\bibitem[Jarrett et al.(2011)]{jarrett11} Jarrett, T. H., Cohen, M., Masci, F., Wright, E., et al.\ 2011, \apj, 735, 112
\bibitem[Juneau et al.(2013)]{juneau13} Juneau, S., Dickinson, M., Bournaud, F., Alexander, D. M., et al.\ 2013, \apj, 764, 176 
\bibitem[Kauffmann et al.(2003a)]{kauffmann03a} Kauffmann, G., Heckman, T. M., Tremonti, C., Brinchmann, J., et al.\ 2003, \mnras, 346, 1055
\bibitem[Kauffmann et al.(2003b)]{kauffmann03b} Kauffmann, G., Heckman, T. M., White, S. D. M., Charlot, S., et al.\ 2003, \mnras, 341, 33
\bibitem[Kauffmann et al.(2003c)]{kauffmann03c} Kauffmann, G., Heckman, T. M., White, S. D. M., Charlot, S., et al.\ 2003, \mnras, 341, 54
\bibitem[Kellermann et al.(1989)]{kellermann89} Kellermann, K. I., Sramek, R., Schmidt, M., Shaffer, D. B. \& Green, R.\ 1989, \aj, 98, 1195
\bibitem[Kennicutt et al.(1989)]{kennicutt89} Kennicutt, R. C., Edgar, B. K., \& Hodge, P. W.\ 1989, \apj, 337, 761
\bibitem[Kennicutt et al.(1994)]{kennicutt94} Kennicutt, R. C., Jr., Tamblyn, P., \& Congdon, C. E.\ 1994, \apj, 435, 22
\bibitem[Kennicutt (1998a)]{kennicutt98a} Kennicutt, R. C., Jr.\ 1998, \araa, 36, 189
\bibitem[Kennicutt (1998b)]{kennicutt98b} Kennicutt, R. C., Jr.\ 1998, \apj, 498, 541
\bibitem[Kennicutt \& Evans (2012)]{kennicutt12} Kennicutt, R. C., Jr. \& Evans, N. J.\ 2012, \araa, 50, 531
\bibitem[Kewley et al.(2001)]{kewley01} Kewley, L. J., Heisler, C. A., Dopita, M. A., \&  Lumsden, S.\ 2001, \apjs, 132, 37
\bibitem[Kewley et al.(2006)]{kewley06} Kewley, L. J., Groves, B., Kauffmann, G., \& Heckman, T. M.\ 2006, \mnras, 372, 961 
\bibitem[Komugi et al.(2007)]{komugi07} Komugi, S., Kohno, K., Tosaki, T., Nakanishi, H., et al.\ 2007, \pasj, 59, 55
\bibitem[K{\"o}nig et al.(2009)]{konig09} K{\"o}nig, S., Eckart, A., Garc{\'i}a-Mar{\'i}n, M., \& Huchtmeier, W. K.\ 2009, \aap, 507, 757
\bibitem[Kormendy \& Ho (2013)]{kormendy13} Kormendy J. \& Ho L. C.\ 2013, \araa, 51, 511
\bibitem[Lacy et al.(2004)]{lacy04} Lacy, M., Storrie-Lombardi, L. J., Sajina, A., Appleton, P. N., et al.\ 2004, \apjs, 154, 166
\bibitem[Laurent et al.(2000)]{laurent00} Laurent, O., Mirabel, I. F., Charmandaris, V., Gallais, P., et al.\ 2000, \aap, 359, 887 

\bibitem[Maragkoudakis et al.(2014)]{maragkoudakis14} Maragkoudakis, A., Zezas, A., Ashby, M. L. N. \& Willner, S. P.\ 2014, in press (astro-ph 1404.0620) 
\bibitem[Martin et al.(2010)]{martin10} Martin, A. M., Papastergis, E., Giovanelli, R., Haynes, M. P., et al.\ 2010, \apj, 723, 1359 
\bibitem[Martin et al.(2012)]{martin12} Martin, A. M., Giovanelli, R., Haynes, M. P., \& Guzzo, L.\ 2012, \apj, 750, 38
\bibitem[Magorrian et al.(1998)]{magorrian98} Magorrian, J., Tremaine, S., Richstone, D., et al.\ 1998, \aj, 115, 2285 
\bibitem[McKee \&  Ostriker (2007)]{mckee07} McKee, C. F. \& Ostriker, E. C.\ 2007, \araa, 45, 565
\bibitem[Mendez et al.(2011)]{mendez11} Mendez, A. J., Coil, A. L., Lotz, J., Salim, S., et al.\ 2011, \apj, 736, 110
\bibitem[Meyer et al.(2004)]{meyer04} Meyer, M. J., Zwaan, M. A., Webster, R. L., Staveley-Smith, L., et al.\ 2004, \mnras, 350, 1195
\bibitem[Mirabel \& Wilson (1984)]{mirabel84} Mirabel, I. F. \& Wilson, A. S.\ 1984. \apj, 277, 92
\bibitem[Moran et al.(2002)]{moran02} Moran, E. C., Filippenko, A. V. \& Chornock, R.\ 2002, \apj, 579, 71
\bibitem[Nagar et al.(2005)]{nagar05} Nagar, N. M., Falcke, H., \& Wilson, A. S.\ 2005, \aap, 435, 521
\bibitem[Pineau et al.(2011)]{pineau11} Pineau, F.-X., Motch, C., Carrera, F., Della Ceca, R., et al.\ 2011, \aap, 527, 126
\bibitem[Pracy et al.(2014)]{pracy14} Pracy, M. B., Owers, M. S., Zwaan, M., Couch, W., et al.\ 2014, \mnras, 443, 388 
\bibitem[Risaliti et al.(1999)]{risaliti99} Risaliti, G., Maiolino, R. \& Salvati, M.\ 1999, \apj, 522, 157
\bibitem[Salim et al.(2007)]{salim07} Salim, S., Rich, R. M., Charlot, S., Brinchmann, J., et al.\ 2007, \apjs, 173, 267
\bibitem[Sanders \& Mirabel (1996)]{sanders96} Sanders, D. B. \& Mirabel, I. F.\ 1996, \araa, 34, 749
\bibitem[Sarzi et al.(2010)]{sarzi10} Sarzi, M., Shields, J. C., Schawinski, K., Jeong, H., et al.\ 2010, \mnras, 402, 2187
\bibitem[Schmidt (1959)]{schmidt59} Schmidt, M.\ 1959, \apj, 129, 243
\bibitem[Seyfert (1943)]{seyfert43} Seyfert, C. K.\ 1943, \apj, 97, 28
\bibitem[Singh et al.(2013)]{singh13} Singh, R., van de Ven, G., Jahnke, K., Lyubenova, M., et al.\ 2013, \aap, 558, 43
\bibitem[Smith et al.(2012)]{smith12} Smith, R. J., Lucey, J. R., Price, J., Hudson, M. J. \& Phillipps, S.\ 2012, \mnras, 419, 3167
\bibitem[Spoon et al.(2007)]{spoon07} Spoon, H. W. W., Marshall, J. A., Houck, J. R., Elitzur, M, et al.\ 2007, \apj, 654L, 49
\bibitem[Springob et al.(2005)]{springob05} Springob, C. M., Haynes, M. P., Giovanelli, R., \& Kent, B. R.\ 2005, \apjs, 160, 149
\bibitem[Stasi{\'n}ska et al.(2008)]{stasinska08} Stasi{\'n}ska, G., Vale Asari, N., Cid Fernandes, R., Gomes, J. M., et al.\ 2008, \mnras, 391, L29
\bibitem[Stern et al.(2005)]{stern05} Stern, D., Eisenhardt, P., Gorjian, V., Kochanek, C. S., et al.\ 2005, \apj, 631, 163
\bibitem[Stern et al.(2012)]{stern12} Stern, D., Assef, R. J., Benford, D. J., Blain, A., et al.\ 2012, \apj, 753, 30
\bibitem[Tal et al.(2014)]{tal14} Tal, T., Dekel, A., Oesch, P., Muzzin, A, et al.\ 2014, \apj, 789, 164
\bibitem[Taniguchi et al.(2000)]{taniguchi00} Taniguchi Y., Shioya Y., \& Murayama T.\ 2000, \aj, 120, 1265
\bibitem[terlevich \& Melnick (1985)]{terlevich85} Terlevich, R. \& Melnick, J.\ 1985, \mnras, 213, 841
\bibitem[Terashima et al.(2000)]{terashima00} Terashima, Y., Ho, L. C., Ptak, A. F., Mushotzky, R. F., et al.\ 2000, \apj, 533, 729
\bibitem[Tremaine et al.(2002)]{tremaine02} Tremaine, S., Gebhardt, K., Bender, R., et al.\ 2002, \apj, 574, 740
\bibitem[Tremonti et al.(2004)]{tremonti04} Tremonti, C. A., Heckman, T. M., Kauffmann, G., Brinchmann, J., et al.\ 2004, \apj, 613, 898
\bibitem[Trinchieri \& di Serego Alighieri (1991)]{trinchieri91}Trinchieri, G. \& di Serego Alighieri, S.\ 1991, \aj, 101, 1647
\bibitem[Toribio et al.(2011a)]{toribio11a} Toribio, M. C., Solanes, J. M., Giovanelli, R., Haynes, M. P. \& Masters, K. L.\ 2011, \apj, 732, 92
\bibitem[Toribio et al.(2011b)]{toribio11b} Toribio, M. C., Solanes, J. M., Giovanelli, R., Haynes, M. P. \& Martin, A. M.\ 2011, \apj, 732, 93
\bibitem[Tosaki et al.(2011)]{tosaki11} Tosaki, T., Kuno, N., Onodera, S., Miura, R. et al.\ 2011, \pasj, 63, 1171
\bibitem[Tueller et al.(2008)]{tueller08} Tueller, J., Mushotzky, R. F., Barthelmy, S., Cannizzo, J. K., et al.\ 2008, \apj, 681, 113 
\bibitem[Urry \& Padovani (1995)]{urry95} Urry, C. M. \& Padovani, P.\ 1995, \pasp, 107, 803
\bibitem[Veilleux \& Osterbrock (1987)]{veilleux87} Veilleux, S. \& Osterbrock, D. E.\ 1987, \apjs, 63, 295
\bibitem[Waller et al.(1987)]{waller87} Waller, W. H., Clemens, D. P., Sanders, D. B. \& Scoville, N. Z.\ 1987, \apj, 314, 397
\bibitem[Wang \& Wei (2008)]{wang08} Wang, J. \& Wei, J.-Y.\ 2008, \apj, 679, 86
\bibitem[Wang et al.(2009)]{wang09} Wang, J., Wei, J.-Y., \& Xiao, P.-F.\ 2009, \apjl, 693, 66
\bibitem[Wang \& Wei (2010)]{wang10} Wang, J. \& Wei, J.-Y.\ 2010, \apj, 719, 1157
\bibitem[Wang et al.(2013)]{wang13} Wang, J., Zhou, X.-L. \& Wei, J.-Y.\ 2013, \apj, 768, 176
\bibitem[Watson et al.(2009)]{watson09} Watson, M. G., Schr{\"o}der, A. C., Fyfe, D., Page, C. G., et al.\ 2009, \aap, 493, 339
\bibitem[Werner et al.(2004)]{werner04} Werner, M. W., Roellig, T. L., Low, F. J., Rieke, G. H., et al.\ 2004, \apjs, 154, 1
\bibitem[West et al.(2010)]{west10} West, A. A., Garcia-Appadoo, D. A., Dalcanton, J. J., Disney, M. J., et al.\ 2010, \aj, 139, 315
\bibitem[Wright et al.(2010)]{wright10} Wright, E. L., Eisenhardt, P. R. M., Mainzer, A. K., Ressler, M. E., et al.\ 2010, \aj, 140, 1868
\bibitem[Wu et al.(1998)]{wu98} Wu, H., Zou, Z.-L., Xia, X.-Y., \& Deng, Z.-G.\ 1998, \aaps, 132, 181
\bibitem[Wu et al.(2005)]{wu05} Wu, H., Cao, C., Hao, C.-N., Liu, F.-S., Wang, J.-L., Xia, X.-Y., Deng, Z.-G., \& Young, C.~K. S.\ 2005, \apjl, 632, L79
\bibitem[Wu et al.(2007)]{wu07} Wu, H., Zhu, Y.-N., Cao, C., \& Qin, B.\ 2007, \apj, 668, 87
\bibitem[Woo et al.(2010)]{woo10} Woo, J. H., Treu, T., Barth, A. J., et al. 2010, \apj, 716, 269
\bibitem[Yan \& Blanton (2012)]{yan12} Yan, R., \& Blanton, M. R.\ 2012, \apj, 747, 61
\bibitem[York et al.(2000)]{york00} York, D. G., et al.\ 2000, \aj, 120, 1579 
\bibitem[Zehaviet al.(2011)]{zehavi11} Zehavi, I., Zheng, Z., Weinberg, D. H., Blanton, M. R., et al.\ 2011, \apj, 736, 59
\bibitem[Zhu et al.(2008)]{zhu08} Zhu, Y.-N., Wu, H., Cao, C., \& Li, H.-N.\ 2008, \apj, 686, 155
\bibitem[Zhu et al.(2010)]{zhu10} Zhu, Y.-N., Wu, H., Li, H.-N., \& Cao, C.\ 2010, Research in Astronomy and Astrophysics, 10, 329
\bibitem[Zinnecker \& Yorke (2007)]{zinnecker07} Zinnecker, H. \& Yorke, H. W.\ 2007, \araa, 45, 565
\end{thebibliography}
\end{document}